\documentclass[final,5p,times,twocolumn,lefttitle]{elsarticle}
\usepackage{amssymb}
\usepackage{lipsum}
\usepackage{amsmath}
\usepackage{xcolor}
\usepackage{color}
\usepackage{float} 
\usepackage{graphicx}
\usepackage{soul}
\usepackage[T1]{fontenc} 
\setcitestyle{square} 
\biboptions{comma}
\usepackage[colorlinks=true, linkcolor=red, citecolor=green, urlcolor=magenta]{hyperref}

\newcommand{\gemma}[1]{{\color{red} #1}}
\newcommand{\jc}[1]{{\color{blue} #1}}

\begin{document}
\begin{frontmatter}

\title{Comments on deformed phase-space cosmology, SUSY and black holes.}

\author[First]{O. L\'opez-Aguayo}
\ead{oscar.lopez@ugto.mx}
\author[Second]{J. C. L\'opez-Domínguez}
\ead{jlopez@fisica.uaz.edu.mx}
\author[First]{G. E. P\'erez-Cu\'ellar}
\ead{perezcg2015@licifug.ugto.mx}
\author[First,Third]{M. Sabido}
\ead{msabido@fisica.ugto.mx}
\affiliation[First]{organization={Departamento de F\'sica de la Universidad de Guanajuato},
addressline={A.P. E-143}, city={Le\'on},    
            postcode={37150},
            state={Guanajuato},
            country={M\'exico}} 
\affiliation[Second]{organization={Unidad Acad\'emica de F\'isica, Universidad Aut\'onoma de Zacatecas},
addressline={Calzada Solidaridad esquina con Paseo a la Bufa S/N}, city={Zacatecas},
            postcode={98060}, 
            state={Zacatecas},
            country={ M\'exico}}
\affiliation[Third]{organization={Department of Physics, University of the Basque Country UPV/EHU, 
P.O BOX 644, 48080 Bilbao, Spain.}}
   
\begin{abstract}
This paper presents the construction of the {supersymmetric (SUSY)} deformed 
version of  the time dependent metric that describes the interior of the Schwarzschild BH. First, the  {noncommutative bosonic} model is constructed by inducing the deformation in the minisuperspace coordinates and the associated canonical momentum. Following the methods of SUSY quantum mechanics, the supercharges of the deformed models are constructed and the {deformed} SUSY Wheeler-DeWitt equation is obtained.   The classical dynamics are also obtained using the WKB approximation,  we find the classical solutions and  the conditions to satisfy the Hamiltonian constraint for both the SUSY and bosonic cases. Finally,  the effects of the deformation in the context of deformed phase space cosmology and black holes are discussed.
\end{abstract}

\end{frontmatter}
\section{Introduction}
Black holes have played a fundamental role in probing alternative theories of gravity, including massive gravity, quantum gravity, and related extensions of General Relativity (GR). Among these theories, one of the most influential frameworks is supersymmetry (SUSY), one of the revolutionary ideas in theoretical physics \cite{Wess:1992cp}. This new symmetry relates bosonic and fermionic degrees of freedom, allowing transformations between them and although constructed in the context of particle physics,  the local gauge theory of supersymmetry resulted in supergravity {\cite{Freedman:1976xh}}. After the realization that supergravity is the square root of GR \cite{Teitelboim:1977fs,Tabensky:1977ic} and considering the highly nonlinear character of supergravity,  simplified models where studied. Following these ideas, supersymmetric quantum cosmology was proposed in \cite{octavio}. By obtaining the square root of the Wheeler-DeWitt (WDW) equation, SUSY generalization of cosmological models can be derived. One approach is to construct the SUSY  Hamiltonian of the cosmological model, by defining the ``square root'' of the potential in the minisuperspace variables \cite{graham}. Alternative, a superfield formulation was constructed, this method allowed to introduce matter in the formulation \cite{tkach}. Research in the SUSY version of the WDW equation \cite{death} led to several approaches to SUSY quantum cosmology (for complete and up to date review SUSY quantum cosmology see \cite{moniz_1,moniz_2}).\\ 
{The old idea of a noncommutative {spacetime} was rekindled in the context of super string theory. It was exploited in particle physics, but eventually studied in the context of gravity. Several noncommutative versions of gravity where constructed \cite{Obregon1,ncsdg,Calmet,Wess}. As noncommutative effects are expected to be present near Planck's scale, one can contemplate and inherently noncommutative spacetime at the early ages or near the singularities of black holes. Unfortunately, noncommutative theories of gravity are highly nonlinear making cumbersome the studies of noncommutative gravity. Following the lessons from SUSY cosmology, in \cite{Obregon2} the authors introduce} the effects of noncommutativity  using the methods of nocommutative quantum mechanics \cite{gamboa} on the WDW equation to construct noncommutative quantum cosmology. Moreover,  classical effects of the noncommutative deformation where explored using the WKB approximation of the noncommutative quantum model \cite{eri1,vakili,huicho,shiraishi,rasouli,rasouli2}. 
The initial studies of noncommutativity in black holes considered two fundamentally distinct approaches\footnote{After these early works on noncommutative black holes, other approaches have been developed 
.}. One considered that the effects of noncommutativity eliminated point-like sources for smeared objects \cite{nicolini}, replacing the mass of point particles with a Gaussian distribution and using the corresponding energy-momentum to find spherically symmetric and static solutions to GR. These solutions were interpreted as a noncommutative black hole, from which the thermodynamical properties were derived. The other approach considered is the diffeomorphism between the Schwarzschild and Kantowski-Sachs (KS) metrics. The authors use the methods of noncommutative quantum cosmology on the WDW equation of the KS model to describe the noncommutative Schwarzschild black hole, allowing to study the effects of the noncommutative deformation on the thermodynamical properties of the noncommutative black hole \cite{julio}. 
These two ideas where combined in  \cite{das} where noncommutative supersymmetric quantum mechanics was presented. Moreover, it was applied to cosmology in \cite{yee_susy} where supercharge based approach was constructed for Noncommutative SUSY cosmology. This is achieved by introducing the effects of noncommutativity on the supercharges by using the analogy between the effects of noncommutativity and a magnetic field. Using minimal coupling in the supercharges, one finds the noncommutative SUSY algebra and the corresponding super Hamiltonian.

The main objective of the present work is to investigate a supersymmetric deformed-phase space model of a metric that describes the interior of a Schwarzschild black hole. We exploit the correspondence between the Schwarzschild interior and a time-dependent cosmological metric whose classical Lagrangian reproduces the Schwarzschild interior solution in the commutative limit \cite{Jalalzadeh:2011yp}. We first derive the Wheeler-DeWitt equation for both the commutative and noncommutative minisuperspace models. Subsequently, following the techniques developed in \cite{das,yee_susy}, we construct the corresponding noncommutative supersymmetric Hamiltonian. Finally, we apply the WKB method and obtain the corresponding metric.

The paper is organized as follows, in section 2 we review the bosonic commutative model and and construct the noncommutative version. The SUSY generalization is done in section 3, we construct the supercharges and derive the SUSY WDW equation, we  calculate the classical solutions that  arise from the semiclassical approximation of the SUSY WDW equation obtained from the deformed Hamiltonian and its SUSY version. Finally, section 4 is devoted for discussion and concluding remarks.   
\section{The bosonic model}
One {of} the techniques to explore the interior of the Schwarzschild BH is to use a reparameterization in terms of a time dependent metric. Let us start with the metric 
\begin{equation}\label{KS}
ds^{2}=- \frac{n^{2}\left(t\right)}{\sigma\left(t\right)}dt^{2}+\sigma\left(t\right)dr^{2}+h^{2}\left(t\right)\left(d\vartheta^{2}+\sin^{2}\vartheta d\phi^{2}\right), 
\end{equation}
where $n(t)$ is the lapse function, while $\sigma(t)$ and $h(t)$ are functions of $t$, and play the role of the dynamical variables. It was showed that the classical solutions of such a dynamical system can
be identified with the interior space-time of a Schwarzschild black hole \cite{Jalalzadeh:2011yp}. Using this metric on the Einstein-Hilbert action with the York-Hawking-Gibbons boundary term we get 
\begin{equation}
    \mathcal{L}=-\frac{V'_0}{8 \pi G}\left\{\frac{1}{n(t)}\left[h(t) \dot{h}(t) \dot{\sigma}(t)+\dot{h}^2(t) \sigma(t)\right]-n(t)\right\},
\end{equation}
where $V'_0$ is the integration constant related to the spatial volume angular terms. After the coordinate transformation and lapse rescaling
\begin{align} \label{var}
&\sigma\left(t\right)=\left[\frac{u(t)+v(t)}{u(t)-v(t)}\right]^{2},\quad h\left(t\right)=\frac14\left[u(t)-v(t)\right]^2,\\
&n(t)=\sqrt{\mathcal{N}^2(t) h^{2}(t)\nu(t)},\nonumber
\end{align}
we {obtain} the Lagrangian 
\begin{equation}\label{Lag1}
 \mathcal{L}= - \frac{V_{0}M_{pl}^2}{\mathcal{N}(t)}\left[\dot{u}\left(t\right)^{2}-\dot{v}\left(t\right)^{2}\right]+\frac{V_{0}M_{pl}^2\mathcal{N}(t)}{4}\left[u^2\left(t\right)-v^2\left(t\right)\right],
\end{equation}
where $V'_0= 8\pi V_0$ denotes a rescaled volume.
To construct the Hamiltonian of the model we derive from Eq.(\ref{Lag1}) the canonical momentum, $\pi_u = -(2V_0M_{pl}^2/\mathcal{N})\dot{u}$, $\pi_v =(2V_0M_{pl}^2/\mathcal{N})\dot{v}$ and obtain the  Hamiltonian 

{\begin{equation}\label{Ham1}
H=\frac{\alpha}{2}\left[\pi_v^2(t)-\pi _u^2(t)+V_0^2M_{pl}^4 \left(v(t)^2-u(t)^2\right)\right],
\end{equation}}
{where $\alpha=\frac{\mathcal{N}(t)}{2 V_0 M_{pl}^{2}}$}. 
Also, the momentum $\pi_\mathcal{N} =0$ implies the constraint\footnote{To {simplify} the notation, we won't explicitly write the  temporal dependence of the functions.} 
\begin{equation}\label{constriccion H}
    \mathcal{H} = \frac{1}{4 V_0 M_{pl}^{2}}\left[\pi_v^2(t)-\pi _u^2(t)\right]+\frac{V_0 M_{pl}^2}{4} \left[v^2(t)-u^2(t)\right] \approx 0.
\end{equation}
In these variables the Hamiltonian associated to the action Eq.(\ref{Lag1}) corresponds to a {\it ghost oscillator}, namely the difference of two harmonic oscillators.\\
Several approaches have been considered to incorporate noncommutativity into physical theories. In its introduction to gravitational models, cosmology is one of many examples that has attracted significant interest. There is a broadly explored path with the aim of studying noncommutativity \cite{Obregon2} where the noncommutative deformation is performed in the  minisuperspace variables. Based on this formalism, a noncommutative deformation will be performed on the minisuperspace coordinates and momenta. The procedure we will follow is based on this approach, but the deformation is done in the minisuperpace variables and their associated momentum. We start with the transformation
\begin{equation}\label{nctrans}
\widehat{v}=v+\frac{\theta}{2}\pi_{u},~~~  \widehat{u}=u-\frac{\theta}{2}\pi_{v},~~~
\widehat{\pi}_{v}=\pi_{v}-\frac{\eta}{2}u,~~~ \widehat{\pi}_{u}=\pi_{u}+\frac{\eta}{2}v, 
\end{equation}
where $\{v,u,\pi_v,\pi_u\}$ are the coordinates and momenta in Eq.(\ref{Ham1}) that satisfied the usual Poisson algebra, $\theta$ and {$\eta$} are the constant noncommutative parameters and $\{\widehat{v},\widehat{u},\widehat{\pi}_v,\widehat{\pi}_u\}$ are the new noncommutative minisuperspace coordinates and momenta that satisfy the deformed Poisson a
\begin{equation}\label{algebra}
\{\widehat{u},\widehat{v}\}=\theta,\;\;\;\; \{\widehat{u},\widehat{\pi}_{u}\}=\{\widehat{v},\widehat{\pi}_{v}\}=1+ {\rho},\;\;\;\; \{\widehat{\pi}_u,\widehat{\pi}_v\}=\eta,
\end{equation}
where  $a{\rho}=\theta\eta/4$. It is important to mention, that the deformation in Eq.(\ref{algebra}) is a particular choice. This is an active area of research and alternative relations have been explored  to analyze different physical scenarios, i.e. gravitational collapse \cite{rasouli}, inflation \cite{rasouli2}, among others.
To establish the deformed theory, the starting point is the analogous Hamiltonian to Eq.(\ref{Ham1}) 
but constructed with the variables obeying the modified algebra Eq.(\ref{algebra}). Then, the deformed Hamiltonian is
\begin{align}\label{Ham2_scalar}
H_{\rm NC} &=  \frac{{\alpha}}{2} \left[\left(   (\widehat{\pi}_v^{~2} - \widehat{\pi}_u^{~2}\right) + {\omega}^2\left(\widehat{v}^{~2}-\widehat{u}^{~2}\right)\right]\\
&=\frac{{\gamma}}{2} \left[   (\pi_v^2 - \pi_u^2) + \ell_{\rm NC}^2(v\pi_u+u\pi_v) + {\omega}_{\rm NC}^2(v^2-u^2)    \right], \nonumber
\end{align}
where $\ell_{\rm NC}^2$ and $\omega^2_{\rm NC}$ are constants defined as
\begin{equation}
\ell_{NC}^2 \equiv \frac{\omega^2 \theta - \eta}{1 - \frac{\omega^2 \theta^2}{4}}, ~~~~~{\omega_{NC}}^2 \equiv \frac{\omega^2 - \frac{\eta^2}{4}}{1 - \frac{\omega^2 \theta^2}{4}},
\label{eq:ell_def}
\end{equation}
with $\omega^2 = V_0^2M_{pl}^4$ {denoting} the frequency of the ghost oscillator, and {$\gamma = \alpha(1 - \frac{\omega^2 \theta^2}{4})$}. The crossed term involving products of position and momentum can not be interpreted as an
angular momentum {component}. Also, the deformation {in} Eq.(\ref{nctrans}) represents two nonequivalent physical descriptions, the commutative-frame,{\it``C-frame",}
and the noncommutative-frame, {\it``NC-frame"}. In the {\it``NC-frame"} is the difference of two uncoupled oscillators in deformed space. In the {\it``C-frame"}, we can interpret the deformed model as the difference of two harmonic oscillator in regular phase space, where the effects of  the noncommutative deformation comes into play as a correction of the potential. 
{Next, a WKB-type approximation is applied to the WDW equation $\widehat{H}\Psi = 0$. By utilizing the Hamiltonian in Eq.(\ref{constriccion H}) as an operator, we can verify that the resulting expressions coincide with the classical equations of motion. Consequently, applying this approximation to the noncommutative WDW equation $\widehat{H}_{NC}\Psi = 0$ via Eq.(\ref{Ham2_scalar}) should reliably yield deformed classical equations, which in principle solve the deformed Einstein's field equations.}

\section{The SUSY  deformed phase-space model.}
Several approaches have been suggested to supersymmetrize the WDW-equation for cosmological models. 
The first {models} proposed \cite{octavio}, {emerged} shortly after the appearance of supergravity. They show, that this theory provides a natural  square root of gravity. We follow an alternative method that allows to define a “square root” of the potential in the minisuperspace of the cosmological model of interest, consequently it is possible to find operators whose 
square yields in the Hamiltonian \cite{moniz_1}. One can summarize this approach as an application of the methods of SUSY quantum mechanics 
to quantum cosmology. 
In this approach the Hamiltonian can be written as
\begin{equation}
    2H_0 = G^{\alpha \beta}\pi_{\alpha}\pi_{\beta} + U(q),
\end{equation}
where  {$\left(G^{\mu \nu}\right)={\rm diag}(-1,1)$ is the metric of the minisuperspace}represents the metric in the minisuperspace, {$\pi^{\mu}$ are the bosonic momenta} and the classical potential $U(q^\alpha)$ 
with $(q^\alpha)= (v,u)$ is related to the superpotential $\phi$ by
\begin{equation}
    G^{\alpha \beta} \frac{\partial \phi}{\partial q^{\alpha}}\frac{\partial \phi}{\partial q^{\beta}} = U(q).
\end{equation}
The {supersymmetric} minisuperspace Hamiltonian, $H_S$, becomes 
\begin{equation}
    H_{S} = \frac{{1}}{2}\left(Q\bar{Q} + \bar{Q}Q \right) = H_0 
    + \frac{\partial^2 \phi}{\partial q^{\mu} \partial q^{\nu}}[\bar{\Theta}^{\mu},\Theta^{\nu}],
\end{equation}
where $Q$ and $\bar{Q}$ are the supercharges
\begin{equation}
Q=\Theta^\mu\left( \pi_\mu+i\frac{\partial\phi}{\partial q^\mu} \right), \quad \bar{Q}=\bar{\Theta}^\mu\left( \pi_\mu-i\frac{\partial\phi}{\partial q^\mu} \right),
\end{equation}
and $\Theta^{\mu}$, $\bar{\Theta}^{\mu}$ are Grassmann variables. The supercharges and the Grassmann variables satisfy the superalgebra
\begin{align}
&\left\{Q,\bar{Q}\right\}=2H_{S},\ \ 
\left\{Q,Q\right\}= \left\{\bar{Q},\bar{Q}\right\} =  0,\ \  \left[Q,H_{S}\right] = \left[\bar{Q},H_{S}\right]=0,\nonumber\\
&\left\{\bar{\Theta}^\alpha,\bar{\Theta}^\beta\right\}=\left\{\Theta^\alpha,\Theta^\beta\right\}=0,\quad \left\{\bar{\Theta}^\alpha,\Theta^\beta\right\}=G^{\alpha\beta}.
\end{align}
A particular representation for 
$\Theta^{\mu}$ and $\bar{\Theta}^{\nu}$ can be constructed from a combination of the gamma matrices
\begin{align}
&\Theta^{v} = \frac{1}{2}\left( \gamma^0 + \gamma^3 \right),~~\Theta^{u} = \frac{1}{2}\left( \gamma^1 -i \gamma^2 \right),\\
&\bar{\Theta}^{v} = \frac{1}{2}\left( \gamma^0 -  \gamma^3 \right), 
   ~~~\bar{\Theta}^{u} = \frac{1}{2}\left( \gamma^1 +  i\gamma^2 \right),\nonumber
\end{align}
where $\gamma^{\mu}$ are 
\begin{equation}
\hspace{-.5cm}
\gamma^{0} = \left(
 \begin{array}{cccc}
 0 & 0 & 0 & -i \\
 0 & 0 & i & 0 \\
 0 & -i & 0 & 0 \\
 i & 0 & 0 & 0 
 \end{array} \right),~~\gamma^{1} = \left(
 \begin{array}{cccc}
 i & 0 & 0 & 0 \\
 0 & -i & 0 & 0 \\
 0 & 0 & i & 0 \\
 0 & 0 & 0 & -i 
 \end{array} \right), 
\end{equation} 
\[
\hspace{-.5cm}
 \gamma^{2} = \left(
 \begin{array}{cccc}
 0 & 0 & 0 & i \\
 0 & 0 & -i & 0 \\
 0 & -i & 0 & 0 \\
 i & 0 & 0 & 0 
 \end{array} \right),~~\gamma^{3} = \left(
 \begin{array}{cccc}
 0 & -i & 0 & 0 \\
 -i & 0 & 0 & 0 \\
 0 & 0 & 0 & -i \\
 0 & 0 & -i & 0 
 \end{array} \right). 
\]
The original construction of noncommutative quantum cosmology \cite{Obregon2}, was based on applying the ideas of noncommutative quantum mechanics \cite{gamboa} to the WDW equation of the Kantowski Sachs cosmological model. Therefore, motivated by this approach, we extend this {approach}rationale here by implementing the framework of noncommutative supersymmetric quantum mechanics \cite{das}.
{Moreover, the effects of the minisuperspace deformation can be interpreted as a commutative theory coupled to a ``magnetic'' field.} 
Consequently for the {noncommutative} deformed SUSY model, we will follow the usual SUSY quantum mechanics and introduce the full deformation 
(coordinates and momenta) as minimal coupling. This gives the same deformed Hamiltonian Eq.(\ref{Ham2_scalar}) that we 
get from using the deformed algebra Eq.(\ref{algebra}), but with a different frequency for the potential.
From  the potential of Eq.(\ref{Ham1}) we calculate the superpotential $\phi$, then the resulting supercharges are
\begin{align}
&Q=\Theta^v\left(\pi_v+\frac{\omega_A^2}{2} u+i \omega v\right)+\Theta^u\left(\pi_u-\frac{\omega_A^2}{2} v-i \omega  u\right),\\
&\bar{Q}=\bar{\Theta}^v\left(\pi_v+\frac{\omega_A^2}{2} u-i \omega  v\right)+\bar{\Theta}^u\left(\pi_u-\frac{\omega_A^2}{2} v+i \omega  u\right),\nonumber
\end{align}
where the deformation is introduced by using the vector potential {${A}_{v}=-\frac{\omega_A^2}{2}{u}$ and ${A}_{u}=\frac{\omega_A^2}{2}{v}${, with a constant $\omega_A$}}, and $\omega$ is the frequency of the {ghost ``oscillators''}. The Hamiltonian operator (after diagonalization) becomes
\begin{gather}
 H_{\rm SNC}=\begin{bmatrix} H_{NC}-i\frac{\ell_{k s}^2}{2} & 0 & 0 & 0 \\ 
 0 & H_{NC}+i\frac{\ell_{k s}^2}{2} & 0 & 0\\
 0 & 0 & H_{NC}-\omega & 0\\
 0 & 0 & 0 & H_{NC}+\omega
 \end{bmatrix},
\end{gather}
where  $H_{NC}=\frac{1}{2}\left[\pi_v^2-\pi_u^2+\ell_{k s}^2\left(v \pi_u+u \pi_v\right)+\omega_{k s}^2\left(v^2-u^2\right)\right]$. In a more compact form
\begin{equation}\label{HSUSYNC}
H_{\rm SNC}= H_{NC}~ \mathbb{I}+\frac{\omega_A^2}{2} \mathbb{A}+\omega \mathbb{B},
\end{equation} 
where $\mathbb{I}$ is the $4\times4$ identity, $\ell_{k s}^2\equiv \omega_A^2$, $\omega_{k s}^2\equiv\left(\omega^2-\omega_A^4 / 4\right)$, $\mathbb{A}=\operatorname{diag}(-i, i, 0,0)$ and $\mathbb{B}=\operatorname{diag}(0,0,-1,1)$.
It is worth noting that the SUSY deformed Hamiltonian Eq.(\ref{HSUSYNC}) shares the same functional form as the noncommutative Hamiltonian Eq.(\ref{Ham2_scalar}), differing only by additive constants. {If we calculate the supercharges for the commutative Hamiltonian, we get the same result as simply taking {$\ell_{ks} =0$} in the SUSY deformed Hamiltonian.
 If  write the Hamiltonian constraint as $H=H_{NC}+k$, two of the components are for {the additive constant} { $k=\pm i l_{ks}^2$ } {and} the last two for $k={\pm}\omega$.} 
{To derive the deformed SUSY WDW equation, we proceed as in \cite{julio_octavio} and apply the canonical quantization to the deformed Hamiltonian given in Eq.(\ref{HSUSYNC})}
After diagonalization of the Hamiltonian and acting on the wave function $\Psi = (\psi_1,\psi_2,\psi_3,\psi_4)$, one gets four equations these are
\begin{align}\label {4H}
&\left(H_{NC} -\frac{i\omega_A^2}{2}\right)\psi_1 = 0,~~ \left(H_{NC} +\frac{i\omega_A^2}{2}\right)\psi_2 = 0,\\
&\left(H_{NC} -\omega^2\right)\psi_3 = 0,~~ 
\left(H_{NC} + \omega^2\right)\psi_4 = 0, \nonumber
\end{align}
{where $H_{\rm NC}$ is given by Eq.(\ref{Ham2_scalar})}. These  four equations can be written as
\begin{equation}\label{NCSUSYWDW}
    \left[ \left(\frac{\partial^2}{\partial u^2} - \frac{\partial^2}{\partial v^2}\right) + i \ell_{ks}^2\left(v\frac{\partial}{\partial u} + u\frac{\partial}{\partial v}\right) +\omega^2_{ks} (u^2 - v^2)\right]\psi_i = k\psi_i,
\end{equation}
where the constant $k$ takes the values $\pm i l^2_{ks},\pm \omega$. {Although the deformed SUSY WDW equation Eq.(\ref{NCSUSYWDW}) resembles the difference of two harmonic oscillator Hamiltonians, the presence of an additional term that is not associated with angular momentum does not admit a straightforward solution}, as the Hamiltonian is the difference of two harmonic oscillators\footnote{Unlike in \cite{yee_susy}, where this idea was explored in phantom cosmology, we were able to interpret the effects of the deformation as an additional angular momentum term}.  
To find the classical solutions we follow the ideas in \cite{julio_octavio}, we simply take  the WKB approximation on the {deformed} SUSY WDW equation. For the WKB approximation we take a wave function of the form $\psi = \exp{\left[\frac{i}{\hbar} \left(S_{1}(v)+ S_{2}(u)\right)\right]}$. Using the approximation 
\begin{equation}
\left( \frac{\partial S_1(v)}{\partial v} \right)^2 > > \frac{\partial^2 S_1(v)}{\partial v^2},\quad \left( \frac{\partial S_2(u)}{\partial u} \right)^2 > > \frac{\partial^2 S_2(u)}{\partial u^2},
\end{equation}
on Eq.(\ref{4H}), we get the corresponding Hamilton-Jacobi equation
\begin{align}
&\left( \frac{\partial S_1(v)}{\partial v} \right)^2 - \left( \frac{\partial S_2(u)}{\partial u} \right)^2 + {\omega}_{ks}^2(v^2 - u^2)\\
&+ \ell_{ks}^2 \left( \frac{\partial S_2(u)}{\partial u} \right)v + \ell_{ks}^2 \left( \frac{\partial S_1(v)}{\partial v} \right)u + 2k = 0,\nonumber
\end{align}
and by identifying  $\frac{\partial S_1(v)}{\partial v} = \pi_v$, $\frac{\partial S_2(u)}{\partial u} = \pi_u$  we get
\begin{equation}
 \hspace{-.5cm}
 2H=\pi_v^2 - \pi_u^2 + \ell^2_{ks}\left(v\pi_u+u\pi_v\right)
 +\omega_{ks}^2\left( v^2-u^2\right) +{ 2k}. 
\end{equation}
{Consequently, at the classical level we arrive at a similar Hamiltonian as in Eq.$(\ref{Ham2_scalar})$ but with a modified frequency} {$\omega^2_{ks}$}. 


Now that we have constructed the deformed and SUSY deformed version and have the corresponding classical limits,  we can analyze the effects of these modifications.\\
Let us start with the commutative case, in \cite{Jalalzadeh:2011yp} using the Hamiltonian constraint in Eq.(\ref{constriccion H}) the authors show that the model describes the interior of a BH. 
They calculate the equations of motion, with the solutions  given by
\begin{equation}\label{eq:uvsols}
    u=A \cos\left(M_{p l} t+\theta_1\right), \quad v=B\cos\left(M_{p l }t+\theta_2\right),
\end{equation}
{where $A,~B,~\theta_1$ and $\theta_2$ are integration constants. Upon} Inserting the solution to the Hamiltonian constraint, establishes a relationship between the constants, such that it is satisfied if $A=\pm B$. We also perform the time rescaling ${n}dt \rightarrow d\tau$, where ${n} = (M_{pl}/2)(u^2 - v^2)$
\begin{equation}
    d\tau =-\frac{M_{pl}}{2} A^2 \sin\left(\Delta \theta\right) \sin\left(2 M_{p l} t+\theta_1+\theta_2\right) d t
\end{equation}
{where we set  $\mathcal{N} = M_{pl}/2$} and $\Delta\theta = \theta_1 - \theta_2$. After integrating the previous equation, we get
\begin{equation}
    \tau-\tau_0=\frac{A^2}{4}\sin\left(\Delta\theta\right)\cos \left(2 M_{p l} t+\theta_1+\theta_2\right).
    \label{eq:time_rescaling}
\end{equation}
Now we write the mass in terms of the dynamical variables, by first matching the coordinates with the components of the Schwarzschild metric (with the interchange  $t\leftrightarrow r$).
Moreover, from the definition of the coordinates in Eq.(\ref{var}) and writing $(\dot{u}, \dot{v})$ in terms of  ($u,v$), we can express the mass in terms of 
$v,~u,~ \pi_v\text{ and } \pi_u$. Finally, after
taking $n = 1$, $G = M_{pl}^{-2}$ and substituting the solutions Eq.($\ref{eq:uvsols}$)
we can  reconstruct the minisuperspace variables $n(t),~\sigma(t)\text{ and } h(t)$. 
After fixing the constant 
{$\tau_0=-\frac{A^2}{4} \sin (\Delta \theta)$ } we get
\begin{align}
    &{h(\tau)}=-\tau \tan \left(\frac{\Delta\theta}{2}\right),\\ 
    &{\sigma(\tau)}=\cot ^2 \left(\frac{\Delta\theta}{2}\right)\left(-\frac{2 M G \cot \left(\frac{\Delta\theta}{2}\right)}{\tau} - 1\right),\nonumber
\end{align}
and by taking $\Delta \theta=-\frac{\pi}{2}+2 n \pi$ one recovers
\footnote{Following the same procedure we cover the same result if we take $B=-A$, the expressions for $h$ and $\sigma$ are the same but interchanging $\cot{\delta}$ with $\tan{\delta}$ and after fixing
 $\Delta \theta=2 \delta=\frac{\pi}{2}+2 n \pi$ one also recovers 
 Eq.(\ref{KS}). } 
  \begin{equation}
ds^{2}=-\left(\frac{2m}{t}-1\right)^{-1}dt^{2}+\left( 1-\frac{2m}
{t}\right)dr^{2}+t^{2}\left(d\vartheta^{2}+\sin^{2}\vartheta d\varphi^{2}\right).
\end{equation} 
{Consequently, the authors in \cite{Jalalzadeh:2011yp} show that the classical solutions to this system correspond to the interior spacetime of a black hole.} 
In order to understand the effects the minisuperspace deformation on the model, we follow the same approach,  {using} the relationship between the cosmological and the Schwarzschild black hole models. As the phase space deformation was applied on the minisuperspace variables associated to the interior of the Schwarzschild black hole, by following {the same}  procedure we  reconstruct the {metric} for the deformed bosonic and SUSY models. Mainly, we find the classical solutions, verify {that they} satisfy the deformed Hamiltonian constraint and reconstruct the metric.\\
Starting form the deformed Hamiltonian 
 Eq.(\ref{HSUSYNC}), The equations of motion are
\begin{equation}
\begin{array}{ll}
\dot{v}=\pi_v+\frac{\ell_{ks}^2}{2} u, & \dot{\pi}_v=-\frac{\ell_{ks}^2}{2} \pi_u-\omega_{ks}^2 v, \\
\dot{u}=-\pi_u+\frac{\ell_{ks}^2}{2} v & \dot{\pi}_u=-\frac{\ell_{ks}^2}{2} \pi_u+\omega_{ks}^2 u.
\end{array}
\end{equation}
We construct a system of coupled second order differential equations {which} can be decoupled with the change of variables $v = \frac{1}{2}(\eta - \rho)$ and $u = \frac{1}{2}(\eta + \rho)$, 
\begin{equation}
\begin{aligned}
& \frac12 \ddot{\eta}- \frac{\ell^2_{ks}}{2}\dot{\eta}+\left(\frac{\ell_{ks}^4}{8} +\frac{\omega_{ks}^2}{2}\right) \eta=0, \\
& \frac12 \ddot{\rho}+ \frac{\ell^2_{ks}}{2}\dot{\rho}+\left(\frac{\ell_{ks}^4}{8} +\frac{\omega_{ks}^2}{2}\right) \rho=0,
\end{aligned}
\end{equation}
solving for $\eta$ and $\rho$ we recover the solutions for $u$ and $v$
\begin{equation}
\begin{aligned}
v&=\frac{1}{2} \left[e^{(\ell_{ks}^2/ 2) t}\left(A e^{i \omega_{ks} t}+B e^{-i \omega_{ks} t}\right)-e^{-(\ell_{ks}^2 / 2) t}\left(C e^{i \omega_{ks} t}+D e^{-i \omega_{ks} t}\right)\right],\\
u&=\frac{1}{2} \left[e^{(\ell_{ks}^2/ 2) t}\left(A e^{i \omega_{ks} t}+B e^{-i \omega_{ks} t}\right)+e^{-(\ell_{ks}^2 / 2) t}\left(C e^{i \omega_{ks} t}+D e^{-i \omega_{ks} t}\right)\right],
\end{aligned}
\label{eq:complete_sols}
\end{equation}
where $A,B,C\text{ and } D$ are constants. 
These solutions must satisfy the Hamiltonian constraint 
\begin{equation}
    \pi_v^2-\pi_u^2+\ell_{k s}^2\left(v \pi_u+u \pi_v\right)+\omega_{k s}^2\left(v^2-u^2\right)+k \approx 0.
\end{equation}
This gives a relation between the constants $A,B,C\text{ and } D$
\begin{equation}
    k=2 \omega_{k s}^2(B C+A D) + i \ell_{k s}^2 \omega_{k s}(B C-A D),
\end{equation}
where $k = \{-i\ell_{ks}^2, i\ell_{ks}^2, -2\omega, 2\omega\}$ and $k=0$ for the bosonic case. Remember that $\ell_{ks}^2 = \omega_A^2$, is the frequency of the vector potential in the deformation of the super charges, $\omega_{k s}^2=\left(\omega^2-\omega_A^4/4\right)$ is the frequency-like coefficient of the modified Hamiltonian and $\omega$ is the frequency of the original ghost oscillator. 
Moreover, the solutions after satisfying the Hamiltonian constraint will be used to reconstruct the metric.
For the cases $k= \pm2\omega$, the constants in the classical solutions are fixed as 
\begin{equation}
    D=\frac{\pm \omega}{2 A \omega_{k s}^2} \quad \text { and } \quad C=\frac{ \pm \omega}{2 B \omega_{k s}^2},
\end{equation}
then, as in the commutative case we perform the time rescaling ${n}dt \rightarrow d\tau$, where ${n} = \pm \frac{\omega M_{pl}}{4 A B \omega_{k s}^2} e^{-2 i \omega_{k s} t}\left(B+A e^{2 i \omega_{k s} t}\right)^2 $, such that the coordinates are
\begin{equation}
\begin{aligned}
& {h(t)}=\frac{\omega^2}{16 A^2 B^2 \omega_{k s}^4} e^{-\left(\ell_{k s}^2+2 i \omega_{k s}\right)t}\left(B+A e^{2 i \omega_{k s} t}\right)^2=\tau, \\
& {\sigma(t)}=\frac{4 A^2 B^2 \omega_{k s}^4}{\omega^2} e^{2 \ell_{k s}^2 t}=f(\tau).
\end{aligned}
\end{equation}
After direct substitution between the above equations 
\begin{equation}
    \tau=\frac{ \pm \omega e^{-\ell_{ks} t }}{4 A B \omega_{ks}^2 M_{pl}},\quad f(\tau) = \frac{1}{4 M_{pl}^2 \tau^2}.
\end{equation}
we get the metric 
\begin{equation}
    ds^2=-f^{-1}(\tau) d \tau^2+f(\tau) d r^2+\tau^2 d \Omega^2.
    \label{eq:line_element}
\end{equation}
It is important to point out that when doing the reparameterization, the effects of the deformation parameters are absorbed in the the time rescaling and cancel out on the construction of the metric. 

For the case $k = \pm i \ell_{ks}^2$, the Hamiltonian constraint fixes the constants in the solution  as 
\begin{equation}
    D=\frac{\pm 1}{2 A \omega_{k s}} \quad \text { and } \quad C=\frac{ \mp 1}{2 B \omega_{k s}},
\end{equation}
as in the previous case we do the  time rescaling ${n}dt \rightarrow d\tau$, with ${n} = e^{-2 i \omega_{k s }t} \frac{M_{p l}}{4 A B \omega_{k s}}\left( \pm B^2 \mp A^2 e^{4 i \omega_{k s} t}\right)$, we get
\begin{equation}   
\begin{aligned}
{h(t)} & =\frac{e^{-\left(\ell_{k s}^2+2 i \omega_{k s}\right) t}\left(B-A e^{2 i \omega_{k s} t}\right)^2}{16 A^2 B^2 \omega_{k s}^2}=\tau, \\
{\sigma(t)} & =\frac{4 A^2 B^2 e^{2 \ell_{ks}^2 t}\left(B+A e^{2 i \omega_{k s} t}\right)^2 \omega_{k s}^2}{\left(B-A e^{2 i \omega_{k s t}}\right)^2}=f(\tau),
\end{aligned}
\end{equation}
and as before, after reconstructing the metric, we get the same result. Then one concludes that the four states of the SUSY deformed Hamiltonian can be written as the line element Eq.$(\ref{eq:line_element})$.\\
For the bosonic deformed case, it is equivalent to solve for $k=0$. The equations of motion do not change with respect to the SUSY case, and we have the same solutions Eq.(\ref{eq:complete_sols}). However, the Hamiltonian constraint does change with respect to the case discussed previously, since in this case the constant $k=0$. 
Therefore, the  Hamiltonian constraint implies the following relation between the constants of the solutions
\begin{equation}
    i \ell_{NC}^2(B C-A D)+2 \omega_{N C}(B C+A D)=0.
\end{equation}
To satisfy this condition, the real and imaginary part must be zero, then $AD = 0$ and $BC=0$. This gives several cases. For $A = B = 0$ is the trivial solution, and although the constraint is satisfied the metric variables are not defined. The only conditions that satisfy the constraint and lead to defined metric variables are $A=C=0$ (with $B\ne0$ and $D\ne0$)
or $B=D=0$ (with $A\ne0$ and $C\ne0$).
For the $A=C=0$, after the  time rescaling ${n}dt \rightarrow d\tau$, with ${n} = \frac{B D M_{pl}}{2} e^{-2 i \omega_{NC} t}$
we get
\begin{equation}
{h(t)}=\frac{D^2}{4} e^{-\left(l_{NC}^2+2 i \omega_{NC}\right)}=\tau, \quad{\sigma(t)}=\frac{B^2}{D^2} e^{2 \ell_{NC}^2 t}=f(\tau),
\end{equation}
and after straightforward substitution, fixing $B=D$ we get 
    \begin{equation}
        \tau=\frac{ e^{-\ell_{NC}^2t}}{2M_{pl}}, \qquad f(\tau) = \frac{1}{4 M_{pl}^2 \tau^2}.
    \end{equation}
For the other possibility $B=D=0$, for the time rescaling ${n}dt \rightarrow d\tau$ with ${n}=\frac{AC M_{pl}}{2} e^{2 i \omega_{NC} t}$, we get
the same result as the previous case. Finally, after reconstructing the metric, we see that the two bosonic cases give the same which is a singular spacetime as in the SUSY model Eq.$(\ref{eq:line_element})$.  

\section{Final Remarks}

In this paper we have constructed the deformed phase space version and SUSY deformed phase space version of the  the time dependent metric proposed in \cite{Jalalzadeh:2011yp}, that describes the interior of the Schwarzschild BH. First, we construct the commutative deformed model, by inducing the deformation in the minisuperspace coordinates and the associated canonical momentum. Following the methods of SUSY quantum mechanics, we show that if {we} introduce the deformation by considering  minimal coupling on the supercharges, the bosonic part of the resulting Hamiltonian is the same as the one obtained from the techniques of deformed phase {space} cosmology, with a changed natural frequency. This allows to give a physical meaning to the noncommutative deformation, mainly that it arises from a vector potential. After {the} diagonalization of the Hamiltonian constructed from the deformed supercharges, we see that the {effect on the Hamiltonian due to the deformation of the minisuperspace variables is} only  adds a constant on the Hamiltonian constraint. From the deformed phase {space} model, we find the classical solutions and find the conditions on the constants to satisfy the Hamiltonian constraint for both the SUSY and bosonic cases, then the deformed cosmological model is a consistent system. The original approach of noncommutative cosmology is based on the assumption that introducing the noncommutative deformation, the resulting theory is related to the the cosmological model derived from a fundamental noncommutative theory of gravity. This work is based on the same assumption, therefore one will consider that the deformed models describe the deformed cosmological models. But there is a big difference when one considers the system not as a cosmological model, but as the description of the interior of a noncommutative Black Hole. The changes induced on the Hamiltonian constraint, impose conditions on the constants of the classical solutions that when we reconstruct the metric it does not have the functional form as in the commutative case. As the claim is that the deformed case represents the deformed cosmological model, the interpretation is that of a model with a different interaction. If we want to extend the analogy then one will have to claim that the deformed metric describes the interior of the deformed black hole. The other possibility, is that the relationship between the cosmological and the interior of the black hole is not valid in the deformed theory of gravity. Therefore, we can not claim the deformed (bosonic or SUSY) model represents a deformed black hole.\\
In summary, if we use an appropriate parametrization to model the interior of the Schwarzschild black hole using the spherical symmetric time dependent metric, when introducing the deformation to construct a mathematical consistent model, we can not conclude (at least for this model) that the resulting model represents a deformed phase space SUSY black hole. 
    
\section*{Acknowledgments}
  
{\bf M.S.} is supported by the program ``Estancias Sab\'aticas para la consolidaci\'on de grupos de investigaci\'on'', SECIHTI. {\bf G.E.P-C.} is supported by the program ``Becas Nacionales de Posgrado'', SECIHTI. {\bf J.C.L-D} is supported by the Grant UAZ-2024-39113.
\bibliographystyle{unsrt}
\bibliography{bib}
\end{document}